\documentstyle[12pt]{article}

\begin{document}

\title{Derivative pricing with virtual arbitrage}
\author{
Kirill Ilinski
\thanks{E-mail: kni@th.ph.bham.ac.uk}
 \ and \ Alexander Stepanenko
\thanks{E-mail: ass@th.ph.bham.ac.uk}
\\ [0.3cm]
{\small\it School of Physics and Space Research,
University of Birmingham,} \\
{\small\it Edgbaston B15 2TT, Birmingham, United Kingdom} \\
{\small\it Tel: +44-121-4147323, Fax: +44-121-4144719}
}

\vspace{3cm}

\date{ }

\vspace{1cm}
 
\maketitle

\begin{abstract}
In this paper we derive an effective equation for derivative pricing which
accounts for the presence of virtual arbitrage opportunities and their 
elimination 
by the market. We model the arbitrage return by
a stochastic process and find an equation for the average derivative
price. This is an integro-differential equation which, in the absence of 
the virtual arbitrage or for an infinitely fast market reaction, 
reduces to the Black-Scholes equation. Explicit formulas are obtained for
European call and put vanilla options. 
\end{abstract}

\section{Introduction}
The Black-Scholes (BS) analysis~\cite{BS} of derivative pricing 
is one of the most beautiful results in financial 
economics. The BS formulas are easy to understand and to handle which
leads to wide use of the formulas by traders. There are several assumptions
in the basis of BS analysis such as the quasi-Brownian character of the 
underlying
price process, constant volatility and, the most importantly for our goal, the
absence of arbitrage. Though first two assumptions can be relaxed and
a lot of attempts to improve BS analysis using stochastic volatility models and
price models beyond the quasi-Brownian class have been made, the third 
assumption,
namely the no-arbitrage condition has not been tackled yet. 
Indeed, almost all models
imply the no-arbitrage constraint, employ the martingales and hedging portfolios 
to calculate derivative prices. More generally, the no-arbitrage constraint
is at the heart of contemporary financial calculus~\cite{Duffie,Wilmott} and 
leads to many
fruitful results not only in derivative pricing but also in portfolio theory
and asset pricing~\cite{Ross,portfolio}.

Accepting that the local (virtual) 
arbitrage is short lived and hence irrelevant for long term pricing, one 
is forced to consider the arbitrage if the study of the short term 
behavior is one's objective. It is well-known that local arbitrage 
opportunities are always present in the market~\cite{der} 
and have finite life-time
(for example, for extremely liquid markets such as futures on S\&P the 
characteristic arbitrage time is of the order of 5 minutes~\cite{Sofianos}; 
for less liquid markets such as the bonds market the time can be much longer). 
This poses the question of how to account the local arbitrage opportunities 
and improve the pricing formulas.

One of the possible approaches to the problem has been suggested in 
Ref~\cite{hep-th/9710148}. It was argued there that the violation
of the no-arbitrage constraint and the 
non-Brownian character of underlying price
walks can be accounted for in the same framework.
Being a consistent theory, the approach is however extremely
complicated and does not allow one yet to get simple analytical results which
would be easy understandable and handleable. That is why in this paper we
develop simplified tractable analytical version to account for virtual arbitrage 
opportunities which does not use complicated techniques and results in
simple enough final equations. 
 
The paper is organized as follows. In next section we formulate the model.
To this end the BS equation is rederived and it is shown how the local arbitrage 
opportunities can be introduced in the model. In section 3 we find an equation
for the average derivative price. This
equation generalizes the BS equation and converges 
to it in the limit of absence of the
arbitrage opportunities or infinitely fast market reaction. In conclusion
we discuss drawbacks of the model and possible ways to improve it.
Explicit analytical solutions 
for European vanilla options are contained in Appendix.

\section{Formulation of the model}

We start this section with a quick standard derivation of the BS equation. 
This both makes the paper self-contained, introduces useful notation, and 
allows us to emphasize the important notion which is extensively used later.

First, let us denote $V(t,S)$ as the price of a derivative at time $t$ 
condition to the underlying asset price equal to $S$. We assume that the 
underlying asset price follows the geometrical Brownian motion, i.e.
$$
\frac{{\rm d}S}{S} = \mu {\rm d}t + \sigma {\rm d}W \ ,
$$
with some average return $\mu$ and the volatility $\sigma$. They can be kept
constant or be arbitrary functions of $S$ and $t$. The  symbol ${\rm d}W$ 
stands for the standard Wiener process.

To price the derivative one forms a portfolio which consists of the derivative 
and $\Delta$ units of the underlying asset so that the price of the portfolio is
equal to $\Pi$:
$$
\Pi = V - \Delta S \ .
$$
The change in the portfolio price during a time step $dt$ can be written as
$$
{\rm d}\Pi = {\rm d}V - \Delta {\rm d}S
 = \left(\frac{\partial V}{\partial t} + 
\frac{\sigma^2 S^2}{2}\frac{\partial ^2 V}{\partial S^2}\right) {\rm d}t +
\left(\frac{\partial V}{\partial S} -\Delta\right) {\rm d}S \ , 
$$
from of Ito's lemma.
We can now chose the number of the underlying asset units $\Delta$ to be
equal to $\frac{\partial V}{\partial S}$ to cancel the second term on the right
hand side of the last equation. Since, after cancellation, there are no
risky contributions (i.e. there is no term proportional to ${\rm d}S$) the 
portfolio
is risk-free and hence, in the absence of the arbitrage, its price will grow
with the risk-free interest rate $r$:
\begin{equation}
{\rm d}\Pi = r \Pi{\rm d}t \ ,
\label{p1}
\end{equation}
or, in other words, the price of the derivative $V(t,S)$ shall obey the
Black-Scholes equation:
\begin{equation}
\frac{\partial V}{\partial t} + 
\frac{\sigma^2 S^2}{2}\frac{\partial ^2 V}{\partial S^2}  + r
S\frac{\partial V}{\partial S} - r V = 0 \ .
\label{BS}
\end{equation}
In what follows we use this equation in the following operator form:
\begin{equation}
{\cal L}_{\rm BS} V = 0 \quad , 
\qquad 
{\cal L}_{\rm BS} = \frac{\partial }{\partial t} + 
\frac{\sigma^2 S^2}{2}\frac{\partial ^2 }{\partial S^2}  + r
S\frac{\partial }{\partial S} - r \ .
\label{l}
\end{equation}
To formulate the model we return back to Eqn(\ref{p1}).
Let us imagine that at some moment of time $\tau < t$
a fluctuation of the return (an arbitrage opportunity) appeared in the market. 
It happened when the price of the underlying stock was 
$S^{\prime}\equiv S(\tau)$. We then denote this instantaneous arbitrage return 
as 
$\nu (\tau,S^{\prime})$.
Arbitragers would react to this circumstance and act in such a way that the 
arbitrage gradually disappears and the market returns to its equilibrium state,
i.e. the absence of the arbitrage.
For small enough fluctuations it is natural to assume that the arbitrage return
${\cal R}$ (in absence of other fluctuations) evolves according to the following
equation:
\begin{equation}
\frac{{\rm d}{\cal R}}{{\rm d}t} = - \lambda {\cal R} \ , \qquad 
{\cal R}(\tau) = \nu (\tau,S^{\prime})
\label{R1}
\end{equation}
with some parameter $\lambda$ which is characteristic for the market. 
This parameter can be either estimated from a microscopic theory like
\cite{hep-th/9710148} or can be found from the market using
an analogue of the fluctuation-dissipation theorem~\cite{Landau}. 
In the last case the parameter $\lambda$ can be estimated from the market data 
as
$$
\lambda = -\frac{1}{(t - t^{\prime})}\log \left[
\left\langle\frac{{\cal L}_{\rm BS} V}{V - S\frac{\partial V}{\partial S}}(t)
\frac{{\cal L}_{\rm BS} V}{V - S\frac{\partial V}{\partial S}}(t^{\prime})
\right\rangle_{\rm market} \left/  
\left\langle\left(
\frac{{\cal L}_{\rm BS} V}{V - S\frac{\partial V}{\partial S}}
\right)^{2}(t)\right\rangle_{\rm market}\right.
\right]
$$
and may well be a function of time and the price of the underlying asset. 
In what follows we however consider $\lambda$ as a constant to
get simple analytical formulas for derivative prices. 
The generalization to the case of time-dependent parameters 
is straightforward.

The solution 
of Eqn(\ref{R1}) gives us ${\cal R}(t,S) = \nu(\tau,S) 
{\rm e}^{-\lambda (t-\tau)}$ which,
after summing over all possible fluctuations with the corresponding 
frequencies, leads us to the following expression for the arbitrage return
at time $t$:
\begin{equation}
{\cal R}(t,S) = \int _{0}^{t} {\rm d}\tau \int_{0}^{\infty} 
{\rm d}S^{\prime}\ e^{-\lambda (t-\tau)} 
P(t,S|\tau,S^{\prime})\nu (\tau,S^{\prime}) \ , \qquad t<T
\label{R}
\end{equation}
where $T$ is the expiration date for the derivative contract
started at time $t=0$ and the function $P(t,S|\tau,S^{\prime})$ is the 
conditional probability for the underlying price. To specify the stochastic 
process $\nu(t,S)$ we assume that the fluctuations at different times and 
underlying prices are independent and form the white noise with a variance 
$\Sigma^2 \cdot  f(t)$:
\begin{equation}
\langle\nu(t,S)\rangle =0 \quad \ , \qquad 
\langle\nu(t,S) \nu(t^{\prime}, S^{\prime})\rangle = 
\Sigma^2 \cdot \theta (T-t) f(t) \delta (t-t^{\prime}) \delta (S-S^{\prime}) \ .
\label{nu}
\end{equation}
The function $f(t)$ is introduced here to smooth out the transition to the
zero virtual arbitrage at the expiration date.
The quantity $\Sigma^2 \cdot f(t)$ can be 
estimated from the market data as
$$
\frac{\Sigma^2}{2\lambda} \cdot  f(t) = 
\left\langle\left(
\frac{{\cal L}_{\rm BS} V}{V - S\frac{\partial V}{\partial S}}
\right)^{2}(t)\right\rangle_{\rm market}
$$
and has to vanish as time tends to the expiration date.

Since we introduced the stochastic arbitrage return  ${\cal R}(t,S)$,
Eqn(\ref{p1}) has to be substituted with the following equation:
\begin{equation}
{\rm d}\Pi = [r + {\cal R}(t,S)] \Pi {\rm d} t\ ,
\label{p2}
\end{equation}
which can be rewritten as
\begin{equation}
{\cal L}_{\rm BS} V = {\cal R}(t,S) 
\left(V - S\frac{\partial V}{\partial S}\right) \ ,
\label{a1}
\end{equation}
using the operator ${\cal L}_{\rm BS}$.
Eqns(\ref{a1}),(\ref{R}) and (\ref{nu}) complete the formulation of the model.

It is worth noting that the model reduces to the pure BS analysis in the
case of infinitely fast market reaction, i.e. $\lambda\rightarrow \infty$.
It also returns to the BS model when there are no arbitrage opportunities at 
all, i.e. when $\Sigma =0$.

In the presence of the random arbitrage fluctuations ${\cal R}(t,S)$, the only objects
which can be calculated are the average value and other higher moments of the
derivative price. In this paper we examine the average price and 
derive the pricing equation for it in the next section.

\section{Effective equation for derivative price}

We start this section with the note that the probability distribution of
${\cal R}(t,S)$ is Gaussian. Moreover, it can be shown that the probability 
(up to a normalization constant) of the trajectory ${\cal R}(\cdot ,\cdot )$ 
has the form:
\begin{equation}
P[{\cal R}(\cdot ,\cdot )] \sim \exp\left[ 
 - \frac{1}{2\Sigma^2} \int_0^\infty {\rm d} t 
{\rm d} t^{\prime} {\rm d} S {\rm d} S^{\prime} \
{\cal R}(t,S) K^{-1}(t,S|t^{\prime},S^{\prime}) {\cal R}(t',S')\right] \ ,
\label{pp}
\end{equation}
where the kernel of the operator $K$ is defined as:
\begin{eqnarray}
K(t,S|t^{\prime},S^{\prime}) &=& \theta(T-t)\theta(T-t^{\prime})
\int_0^\infty{\rm d}\tau {\rm d}s\ f(\tau) \theta(t-\tau)\theta(t^{\prime}-\tau)
{\rm e}^{-\lambda(t+t^{\prime}-2\tau)} 
\nonumber\\
&&\times 
P(t,S|\tau,s) P(t^{\prime},S^{\prime}|\tau,s) \ .
\end{eqnarray}
It is easy to see that the kernel is of order $1/\lambda$ and vanishes as
$\lambda \rightarrow \infty$.
Eqn(\ref{pp}), in particular, results in the equality for the correlation 
function:
\begin{equation}
\langle{\cal R}(t,S){\cal R}(t^{\prime},S^{\prime})\rangle = 
\Sigma^2 \cdot K(t,S|t^{\prime},S^{\prime}) \ ,
\label{RR}
\end{equation}
which we use below.

Now let us return to the dynamical equation for the derivative price
\begin{equation}
{\cal L}_{\rm BS}V(t,S) = {\cal R}(t,S)[V(t,S)- S\partial_S V(t,S)] 
\label{a10}
\end{equation}
and note that, since $\Sigma^2 /\lambda$ plays a role of 
small parameter in the problem,
the noise ${\cal R}$ can be considered as weak and we can find a
formal iterative $\cal R$-dependent solution of the last equation. 
In the lowest non-trivial order we have the equation:
\begin{equation}
{\cal L}_{\rm BS}V = {\cal R}[V-S\partial_SV]
 = {\cal R}[1-S\partial_S]{\cal L_{\rm BS}}^{-1}{\cal R}[V-S\partial_S V] \ ,
\label{a11}
\end{equation}
which after averaging (using (\ref{RR})) over all possible realizations of the 
fluctuations $\cal R$ give us an equation for the average derivative 
price $\bar{V}\equiv \langle V\rangle_{\cal R}$ up to and
including terms proportional to $\Sigma^2/\lambda$: 
\begin{equation}
{\cal L}_{\rm BS}\bar V(t,S) = \Sigma^2\int\limits_0^\infty\!
{\rm d}t^{\prime}{\rm d}S^{\prime}\ 
\left[(1-S\partial_S){\cal L}_{\rm BS}^{-1}\right](t,S|t^{\prime},S^{\prime}) 
\cdot K(t,S|t^{\prime},S^{\prime}) 
\left[(1-S\partial_S)\bar V\right](t^{\prime},S^{\prime})
\label{a16}
\end{equation}
together with the payoff condition:
$$
\bar V(T,S) = V_{\rm payoff} (S) \ .
$$

Equation (\ref{a16}) is the central result of the paper. This is 
an integro-differential equation which in the limit $\lambda\rightarrow \infty$
or $\Sigma\rightarrow 0$ reduces to the Black-Scholes equation and 
accounts for local arbitrage opportunities and the corresponding market 
reaction to them. Equations of this type are very familiar in physics 
where they are called one-loop effective Dyson-type equations.

To conclude the section, it is interesting to note that 
due to properties of the integrand
on the right hand side of Eqn(\ref{a16}), the integration is effectively 
limited 
to the interval from time $t$ to the expiration date. It means that any 
mispricings which happened in the past do not influence the derivative price, 
as one would expect, and the only relevant contribution 
comes from future mispricings. The explicit solutions
for European vanilla options are derived in the Appendix.

\section{Conclusion}
In conclusion we want to discuss some obvious drawbacks of the model 
and ways to improve it. 

First of all, all critical comments of BS analysis 
can be forwarded to this model,
except for the no-arbitrage constraint. Indeed 
if there are transaction costs or
if the price process for 
the underlying asset is not quasi-Brownian motion, it is impossible to
create a risk-less portfolio and, hence, to derive the model. 
This is not a new problem and many efforts to overcome this difficulty have
been undertaken. It is possible to demonstrate that the virtual arbitrage 
model can be improved in the same manner by these methods as they succeed
for no-arbitrage BS analysis. This, in principle, allows one to include 
the transaction costs,
the ``fat'' tails for the underlying assets probability distribution function
and the stochastic volatility in
the present model by redefinition of the operator ${\cal L}_{\rm BS}$. 

The second point concerns the market reaction to the arbitrage opportunity, or,
qualitatively the form of  ${\cal R}(t,S)$ in Eqn(\ref{p2}). 
It may be argued that the
market reaction is not exponential as assumed in Eqn(\ref{R}), but has another
functional dependence. This dependence can be found from statistical analysis
of the stock and derivative prices and then included in the equation for
${\cal R}(t,S)$ (in particular, the functional dependence can 
change with time,
for example $\lambda$ in (\ref{R}) can be a function of $\tau$). 
It certainly complicates the model but leaves the general framework
intact.

Another point to consider is the absence of correlations between virtual
arbitrage opportunities which we assumed in the text, i.e. the white noise
character of the process $\nu(t,S)$. It is clear that some correlations
can be easily included in the model by substituting the relations in 
Eqn(\ref{nu}) by the equations:
$$
\langle\nu(t,S)\rangle =0 \ , \qquad 
\langle\nu(t,S) \nu(t^{\prime}, S^{\prime})\rangle = 
\Sigma^2 \cdot F(t,S|t^{\prime},S^{\prime}) 
$$
with some correlation function $F(t,S|t',S')$. Such generalization, though 
making
the analytical study almost impossible, allows one to proceed with numerical 
analysis for the model.

Finally, the model contains new parameters such as $\Sigma$ and $\lambda$.
Though the parameters can be estimated from the statistical 
data as we mentioned above, they still can be considered as a fitting 
parameters. If we follow this line, to define their values some kind of test
of the final formulas on real market data should be carried out. The least 
curve implied volatility can be used as an example of such tests.

\section*{Acknowledgments.}
We want to thank Alexandra Ilinskaia (ANZ IB, London) and Ely Klepfish
(Commerzbank, London) for a number of 
long discussions of the subject. K.I. is very grateful to Yuval 
Gefen (Weizmann Institute, Israel) whose remarks on
possible applications of the fluctuation-dissipation theorem to 
the financial market triggered this work.

\section*{Appendix: Solution for European vanilla options}
Before going further we would like to point out that, since Eqn(\ref{a16}) is 
linear, there exists the call-put parity theorem for the average prices  
$C(t,S,T,E)$ and $P(t,S,T,E)$ of 
European call and put with the same strike price $E$ and the same expiration 
date $T$:
$$
C(t,S,T,E) - P(t,S,T,E) = F(t,S,T,E) \ ,
$$
where $F(t,S,T,E)$ is the price of the corresponding forward agreement. 
It means that pricing of call and put  also gives one a price for the forward 
agreement.

In this appendix we derive explicit formulas for European call and put options
for the case of geometrical random walk for the price of the underlying asset.
In this case it is convenient to use the variable $y\equiv\log(S/E)$ 
instead of $S$. The probability distribution function then takes the form:
$$
P(t,y|t',y') = 
\frac{1}{\sigma\sqrt{2\pi(t-t')}}\exp
\left[
 - \frac{\{y-y'-\mu(t-t')\}^2}{2\sigma^2(t-t')}
\right]
$$
and the operator ${\cal L}_{\rm BS}$ can be rewritten as:
$$
{\cal L}_{\rm BS} = \partial_t + \frac{\sigma^2}{2}\partial_y^2
 + \left(r-\frac{\sigma^2}{2}\right)\partial_y - r \ .
$$
The corresponding kernel $K(t,y|t^{\prime},y^{\prime})$ is equal to
\begin{eqnarray}
K(t,y|t^{\prime},y^{\prime}) 
&=&
\theta(T-t)\theta(T-t^{\prime})\int_0^\infty{\rm d}\tau \ f(\tau)
\theta(t-\tau)\theta(t^{\prime}-\tau)
{\rm e}^{-\lambda(t+t^{\prime}-2\tau)}
\nonumber\\
&&\times
\frac{1}{\sigma\sqrt{2\pi(t+t^{\prime}-2\tau)}}
\exp\left[
 - \frac{\{y-y^{\prime}-\mu(t-t^{\prime})\}^2}{2\sigma^2(t+t^{\prime}-2\tau)}
\right] \ .
\label{aa}
\end{eqnarray}

In the limit of sufficiently fast market relaxation 
($\lambda\gg\sigma^2,\mu$) and the particular choice of the smoothing 
function $f(t)$
\begin{equation}
f(t) = 1 - {\rm e}^{-2\gamma(T-t)}
\label{a7}
\end{equation}
the kernel takes the form
\begin{equation}
K(t,y|t^{\prime},y^{\prime}) = \delta(y-y^{\prime})
\theta(T-t)\theta(T-t^{\prime}) K(t,t^{\prime}) \ ,
\label{a8}
\end{equation}
where the following notation has been introduced
\begin{eqnarray}
K(t,t^{\prime}) &=& 
\frac{1}{2\lambda}
\theta(t-t^{\prime})
\left\{ 
{\rm e}^{-\lambda(t-t^{\prime})}
\left[
1 - \frac{\lambda{\rm e}^{-2\gamma(T-t^{\prime})}}{\lambda+\gamma} 
\right]
 - {\rm e}^{-\lambda(t+t^{\prime})}
\left[
1 - \frac{\lambda{\rm e}^{-2\gamma T}}{\lambda+\gamma} 
\right]
\right\}
\nonumber\\
&+& 
\frac{1}{2\lambda}\theta(t^{\prime}-t) 
\left\{ 
{\rm e}^{-\lambda(t^{\prime}-t)}
\left[
1 - \frac{\lambda{\rm e}^{-2\gamma(T-t)}}{\lambda+\gamma} 
\right]
 - {\rm e}^{-\lambda(t+t^{\prime})}
\left[
1 - \frac{\lambda{\rm e}^{-2\gamma T}}{\lambda+\gamma} 
\right]
\right\}
\label{a9}
\end{eqnarray}
If $\lambda(t+t')\gg1$, that is $t,t'$ are sufficiently far from $t=0$, then 
the terms proportional to ${\rm e}^{-\lambda(t+t^{\prime})}$ in (\ref{a9}) 
can be neglected and we obtain for $K(t,t')$:
\begin{equation}
K(t,t^{\prime}) =
\frac{1}{2\lambda}
\theta(t-t^{\prime})
{\rm e}^{-\lambda(t-t^{\prime})}
\left[
1 - \frac{\lambda{\rm e}^{-2\gamma(T-t^{\prime})}}{\lambda+\gamma} 
\right]
 + \frac{1}{2\lambda}\theta(t^{\prime}-t) 
{\rm e}^{-\lambda(t^{\prime}-t)}
\left[
1 - \frac{\lambda{\rm e}^{-2\gamma(T-t)}}{\lambda+\gamma} 
\right]
\label{a9a}
\end{equation}
A kernel of the integral operator 
${\cal L}_{\rm BS}^{-1}$ in this case can be written as
\begin{equation}
{\cal L}_{\rm BS}^{-1}(t,y|t',y') = 
- \frac{\theta(t'-t)}{\sigma\sqrt{2\pi(t'-t)}}
\exp\Biggl[
 - \frac{\{y'-y-\alpha(t'-t)\}^2}{2\sigma^2(t'-t)} - r(t'-t) 
\Biggr]\ .
\label{a17}
\end{equation}
with the parameter $\alpha$ given by the relation
\begin{equation}
\alpha\equiv r-\frac{\sigma^2}{2} \ .
\label{a17a}
\end{equation}
Now we will solve Eqn(\ref{a16}) keeping in mind that
$\Sigma^2 /\lambda$ is a small parameter in the problem. 
At the first non-trivial order we have:
\begin{equation}
V = V_0 + \frac{\Sigma^2}{\lambda} \cdot V_1
\label{a13}
\end{equation}
where the zero-order solution $V_0(t,y)$ is the BS solution 
for the derivative:
$$
{\cal L}_{\rm BS}\bar V_0 (t,y) = 0 \ , \qquad 
\bar V_0 (T,y) = V_{\rm payoff} (y)
$$
and $\bar{V}_1$ satisfies the equation
\begin{equation}
{\cal L}_{\rm BS}\bar V_1(t,y) = \int\limits_0^\infty\!{\rm d}t'
\int\limits_{-\infty}^\infty\!{\rm d}y'\ 
\left[(1-\partial_y){\cal L}_{\rm BS}^{-1}\right](t,y|t',y') K(t,y|t',y') 
\left[(1-\partial_y)V_0\right](t',y')
\label{a15}
\end{equation}
which can be explicitly rewritten as:
\begin{equation}
{\cal L}_{\rm BS}\bar V_1(t,y) = F(t,y)\ ,\qquad V_1 (T,y) = 0
\label{a20}
\end{equation}
with the right hand side given by the equation:
\begin{eqnarray}
F(t,y) &=& -\frac{1}{4\sigma}\left(1+\frac{2r}{\sigma^2}\right)\theta(T-t)
\left[1 - \frac{\lambda}{\lambda+\gamma}{\rm e}^{-2\gamma(T-t)}\right]
\nonumber\\
&&\times
\int_t^T{\rm d}t'\ 
\frac{{\rm e}^{-(\lambda+r+\alpha^2/2\sigma^2)(t'-t)}}{\sqrt{2\pi(t'-t)}}
\left[(1-\partial_y)V_0\right](t',y) \ .
\label{a21}
\end{eqnarray}

Following these general formulas, the prices of the European call and put
options have the form:
$$
C(t,y) = C_0(t,y) + \frac{\Sigma^2}{\lambda}C_1(t,y) \ ,
\qquad P(t,y) = P_0(t,y) + \frac{\Sigma^2}{\lambda}P_1(t,y) \ .
$$
The zeroth order solutions are just BS solutions for the options:
\begin{eqnarray}
C_0(t,y) &=& 
E\left[{\rm e}^yN[d_1(y,T-t)]-{\rm e}^{-r(T-t)}N[d_2(y,T-t)]\right]
\label{a22}\\
P_0(t,y) &=& 
E\left[{\rm e}^{-r(T-t)}N[-d_2(y,T-t)]-{\rm e}^yN[-d_1(y,T-t)]\right]
\label{a23}
\end{eqnarray}
with the notation
$$
d_1(y,\tau) = \frac{y+(\alpha+\sigma^2)\tau}{\sigma\sqrt{\tau}}\ ,\qquad
d_2(y,\tau) = \frac{y+\alpha\tau}{\sigma\sqrt{\tau}}\ ,
$$
and
$$
N(x)\equiv\frac{1}{\sqrt{2\pi}}\int_{-\infty}^x{\rm d}y\ {\rm e}^{-y^2/2} \ .
$$
The correction terms $C_1(t,y)$, $P_1(t,y)$ can be obtained
after some straightforward algebra:
\begin{eqnarray}
C_1(t,y) &=&  \frac{E{\rm e}^{-r(T-t)}}{4\sigma}
\left(1+\frac{2r}{\sigma^2}\right)
\int\limits_0^{T-t}{\rm d}\tau\ 
\left\{
T-t-\tau
 + \frac{{\rm e}^{-2\gamma(T-t)}-{\rm e}^{-2\gamma\tau}}
   {2\gamma(1+\gamma/\lambda)}
\right\}
\nonumber\\
&&\times
\frac{{\rm e}^{-(\lambda+\alpha^2/2\sigma^2)\tau}}{\sqrt{2\pi\tau}}
N[d_2(y,T-t-\tau)]
\label{a24}\\
P_1(t,y) &=& -\frac{E{\rm e}^{-r(T-t)}}{4\sigma}
\left(1+\frac{2r}{\sigma^2}\right)
\int\limits_0^{T-t}{\rm d}\tau\ 
\left\{
T-t-\tau
 + \frac{{\rm e}^{-2\gamma(T-t)}-{\rm e}^{-2\gamma\tau}}
   {2\gamma(1+\gamma/\lambda)}
\right\}
\nonumber\\
&&\times
\frac{{\rm e}^{-(\lambda+\alpha^2/2\sigma^2)\tau}}{\sqrt{2\pi\tau}}
N[-d_2(y,T-t-\tau)]
\label{a25}
\end{eqnarray}

When the expiration occurs at sufficiently large time that we can neglect
its influence, the smoothing function $f(t)$
can be substituted by 1, i.e. the limit $\gamma\to\infty$ can be taken.
In this case expressions (\ref{a24},\ref{a25}) 
have an even simpler form:
\begin{equation}
C_1(t,y) = \frac{E{\rm e}^{-r(T-t)}}{4\sigma}
\left(1+\frac{2r}{\sigma^2}\right)
\int\limits_0^{T-t}{\rm d}\tau\ (T-t-\tau)
\frac{{\rm e}^{-(\lambda+\alpha^2/2\sigma^2)\tau}}{\sqrt{2\pi\tau}}
N[d_2(y,T-t-\tau)]
\label{a26}
\end{equation}
\begin{equation}
P_1(t,y) = -\frac{E{\rm e}^{-r(T-t)}}{4\sigma}
\left(1+\frac{2r}{\sigma^2}\right)
\int\limits_0^{T-t}{\rm d}\tau\ (T-t-\tau)
\frac{{\rm e}^{-(\lambda+\alpha^2/2\sigma^2)\tau}}{\sqrt{2\pi\tau}}
N[-d_2(y,T-t-\tau)] \ .
\label{a27}
\end{equation}

\end{document}